\documentstyle[11pt]{article}
\begin{document}

\title{Fermion Pair Production From an Electric Field Varying in Two
Dimensions}

\author{J.E. Seger$^*$ \\
Physics Department, Creighton University, \\
Omaha, NE 68178 USA \\
\and
A.B. Balantekin$^{**}$ \\
Physics Department, University of Wisconsin, \\
Madison, Wisconsin, 53706 USA\\}
\date{}

\maketitle

\begin{abstract}
The Hamiltonian describing fermion pair production from an arbitrarily
time-varying electric field in two dimensions is studied using a
group-theoretic approach.  We show that this Hamiltonian can be encompassed
by two, commuting $SU(2)$ algebras, and that the two-dimensional problem can
 therefore be reduced to two one-dimensional problems.  We compare the group
 structure for the two-dimensional problem with that previously derived for
the one-dimensional problem, and verify that the Schwinger result is
obtained under the appropriate conditions.
\end{abstract}

\noindent
Address correspondence to J.E. Seger.

\noindent
E-mail: $^*$jseger@creighton.edu or $^{**}$baha@nucth.physics.wisc.edu

\bigskip

PACS numbers:  03.65.Fd, 12.20.-m, 02.20.Hj, 11.15.Tk
\vfill \eject

\renewcommand{\baselinestretch}{1.5}
\tiny
\normalsize

\section*{I.  INTRODUCTION}
Fermion pair production takes place in a large number of physical
situations; a comprehensive review of its applications in atomic,
nuclear, elementary particle physics, astrophysics and cosmology
is given in Ref.\ \cite{Greiner}. Consequently, the
problem of pair production
from classical external electric fields has been the subject of
considerable theoretical
interest.\cite{Schwinger}-\cite{Balant}  The rate of fermion pair
production from a uniform, static electric field was originally calculated by
Schwinger \cite{Schwinger} to be
\begin{equation}
\omega = \frac{\alpha E^2}{\pi^2}\sum_{n=1}^{\infty}\frac{1}{n^2}
\exp \left( - \frac{n\pi m^2}{|eE|}\right)\quad,
\label{eq:Schwing}
\end{equation}
where m is the mass of
the produced fermions. To date, an analytic formalism that
successfully addresses the general problem of fields which vary
arbitrarily in both time and space has not been developed.
However, numerous approaches have been suggested which address
particular special cases.  We previously discussed an approach for
predicting the rate of pair production from a spatially
homogeneous but arbitrarily time-varying field, provided the
field is constrained to point in a fixed
direction.\cite{Balant}
We now investigate an extension to this formalism that allows for a field
varying
in two dimensions.

We begin by writing the interaction Hamiltonian for fermions in an electric
field.  We adopt the gauge
\begin{equation}
A_0=0\ , A_i=-\int_{-\infty}^t E_i(t)\,dt\quad.
\end{equation}
The interaction Hamiltonian is then given by
\begin{equation}
H_I={-eA_i\int
d^3x\,\overline\psi_{in}(x)\gamma_i\psi_{in}(x)}\quad,
\label{eq:hi}
\end{equation}
where $\gamma_i$ is the $i^{\rm th}$  $4 \times 4$ Dirac
$\gamma$-matrix, and summation over like indices is assumed.  The
incoming Dirac field, $\psi_{in}$, is that of free fermions,
\begin{equation}
\psi_{in}=\int \frac{d^3{\bf k}}{(2\pi)^{3/2}}\sqrt{\frac{m}{k_0}}
\sum_{\beta}\ \bigr[b_{\beta}(k)u_{\beta}(k)e^{-ik\cdot
x}+d^\dagger_{\beta}(\tilde k) v_{\beta}(\tilde k) e^{ik\cdot
x}\bigr]\quad,
\label{eq:dirac}
\end{equation}
where $b$ and $d^{\dagger}$ are the
usual fermion creation and antifermion annihilation operators,
and $u$ and $v$ are the two-component fermion and antifermion
spinors. We use the symbol $k$ to denote the four-vector
$(k_0,{\bf k})$ and $\tilde k$ to denote $(k_0,-{\bf k})$; $k$
represents the initial momentum of the fermions, and is a
time-independent quantity.  The mass of the created fermions is
given by $m$, and their charge by $e$.

We chose the one-dimensional configuration as a starting point
because, if the field varies in only one direction, the
Hamiltonian contains only one Dirac $\gamma$-matrix, and an
$SU(2)$ algebra is sufficient to encompass the Hamiltonian. In
the two-dimensional case, the Hamiltonian contains two Dirac
$\gamma$-matrices.  The appropriate algebra is then an $SO(4)$
algebra, which is isomorphic to two commuting $SU(2)$ algebras, as we
illustrate below.

\section*{II.  PAIR EMISSION FROM A TWO-DIMENSIONAL ELECTRIC FIELD}
For an electric field that varies in the plane defined by the directions
$i=1,2$, the interaction picture Hamiltonian is
\begin{equation}
H = \int d^3{\bf
k}\,\left\lbrace\left(2k_0-2e\frac{A_i k_i}{k_0}\right)\, J_0(k) -
\frac{e\mu_i A_i}{k_0}\left[J_+^{(i)}(k) +
J_-^{(i)}(k)\right]\right\rbrace\quad,
\label{eq:tudham}
\end{equation}
where summation of i
over indices $1$ and $2$ is implied. $\mu_i$ in this expression is
defined as $\mu_i = \sqrt{k_0^2 - k_i^2}$.  We have defined the
operators in analogy
to the one-dimensional case:\cite{Balant}
\begin{eqnarray}
J_+^{(i)}  & =  & \frac{m}{\mu_i}\sum_{\alpha\beta}\
b_{\alpha}^{\dagger}(k)d_{\beta}^{\dagger}(\tilde k) \bar
u_{\alpha}(k) \gamma_i v_{\beta}(\tilde k)\quad,\qquad i=1,2
\nonumber \\
J_-^{(i)}  & =  & \left[J_+^{(i)}\right]^{\dagger}\quad,\qquad
i=1,2\label{jays} \\
J_0 & =  & \frac{1}{2}\,
\sum_{\alpha}\left[b_{\alpha}^{\dagger}(k)b_{\alpha}(k) - d_{\alpha}(\tilde
k)d_{\alpha}^{\dagger}(\tilde k)\right]\quad.\nonumber
\end{eqnarray}
With an additional operator,
\begin{eqnarray}
Q & = & \sum_{\alpha\beta}\
\left[b_{\alpha}^{\dagger}b_ {\beta} \bar
u_{\alpha}\gamma_3\gamma_5u_{\beta} + d_{\alpha}d_{\beta}^{\dagger}\bar
v_{\alpha}\gamma_3\gamma_5v_{\beta} \right]\quad,\label{cue}
\end{eqnarray}
these operators form an $SO(4)$ algebra.

 From linear combinations of these
operators, we can form two commuting $SU(2)$ algebras, which we denote
by $I_{+,-,0}$ and $T_{+,-,0}$, as follows:
\begin{eqnarray}
I_+ & =  & a\,J_+^{(1)} + a^*\,J_+^{(2)} \quad, \nonumber \\
I_- & =  & \left[I_+\right]^{\dagger}\quad, \label{eyes}\\
I_0 & =  & \frac{1}{2}\,J_0 + \frac{m}{4\mu}\,Q \quad, \nonumber \\
T_+ & = & a^*\,J_+^{(1)} + a\,J_+^{(2)} \quad, \nonumber \\
T_- & = & \left[T_+\right]^{\dagger} \quad,\label{tees}\\
 {\rm and}\qquad T_0 & = & \frac{1}{2}\,J_0 -
\frac{m}{4\mu}\,Q \quad, \nonumber
\end{eqnarray}
where
\begin{equation}
a = \sqrt{\frac{\mu_1\mu_2}{8(\mu_1\mu_2\, - \, k_1k_2)}}\,
+\, i\sqrt{\frac{\mu_1\mu_2}{8(\mu_1\mu_2\, + \, k_1k_2)}} \quad,
\end{equation}
and we have
defined $\mu = \sqrt{k_0^2 - k_1^2 - k_2^2}$.
Each of these $SU(2)$ algebras is in the $j=\frac{1}{2}$ representation.

The
group-theoretic approach has been previously discussed by
Perelemov\cite{perelo}, but the algebras we have derived are distinct from the
algebras he considered.
The $SU(2)$ algebras utilized in Ref.\cite{perelo} are constructed from the
Dirac $\gamma$-matrices, whereas the $SU(2)$ commutation relations of the
operators in Eqs.~\ref{eyes} and ~\ref{tees} follow from the completeness and
orthogonality of the Dirac spinors $u(p)$ and $v(p)$.

The Hamiltonian in Eq.\ \ref{eq:tudham} can be rewritten as a linear
combination of
elements of these two $SU(2)$ algebras, and diagonalized via a
Bogoliubov \cite{Bogol} transformation. The Bogoliubov
transformation takes the usual form:
\begin{eqnarray}
\tilde b_{\alpha}(k) & = & {\cal
U}_{\alpha\beta}(k)b_{\beta}(k) + {\cal
V}_{\alpha\beta}(k)d_{\beta}^{\dagger}(\tilde k)\\
\tilde d_{\alpha}(\tilde
k) & = & {\cal X}_{\alpha\beta}(k)d_{\beta}(\tilde k) + {\cal
Y}_{\alpha\beta}(k)b_{\beta}^{\dagger}(k)\quad,
\end{eqnarray}
where the
coefficients are time-dependent, $2\times 2$ matrices. Requiring that the
transformation preserve the canonical commutation relations constrains the
coefficients to satisfy the relations:
\begin{displaymath}
{\cal UU}^{\dagger} + {\cal
VV}^{\dagger} = 1\quad,\quad{\cal XX}^{\dagger} + {\cal YY}^{\dagger} =
1\quad,
\end{displaymath}
\begin{displaymath}
{\cal UV}^T + {\cal VX}^T = 0\quad.
\end{displaymath}
We further require that the Bogoliubov transformation yield
the diagonal Hamiltonian,
\begin{equation}
H = \int d^s {\bf k}\ \varepsilon(k)
\sum_{\alpha\beta}\,\left[\tilde b_{\alpha}^{\dagger}(k)\tilde b_{\alpha}(k) -
\tilde d_{\alpha}(\tilde k)\tilde d_{\alpha}^{\dagger}(\tilde
k)\right]\quad,
\end{equation}
where $\varepsilon$ is the total energy,
\begin{displaymath}
\varepsilon = \sqrt{m^2 + ({\bf k} - e{\bf A})^2}\quad.
\end{displaymath}
This requirement constrains the coefficients to be:
\begin{eqnarray}
{\cal U} & = & \cos \theta\ I \quad,\\
{\rm and}\qquad {\cal Z} = {\cal U}^{-1}{\cal V}
& = & \frac{-mA_i\tan\theta}{\sqrt{k_0^2{\bf A}^2 -
({\bf A}\cdot{\bf k})^2} } \left[\bar u_{\alpha}\gamma_i v_{\beta}\right]
\quad,
\end{eqnarray}
and summation over $i$ is implied. $\theta$ is
defined by the relations
\begin{equation}
\cos\theta = \sqrt{\frac{\varepsilon + \rho}{2\varepsilon}}\qquad\qquad ,
\qquad\qquad \sin\theta = -\sqrt{\frac{\varepsilon -\rho}{2\varepsilon}}\quad,
\end{equation}
and $\rho$ is defined by
\begin{equation}
\rho = k_0 - \frac{e{\bf A\cdot k}}{k_0}\quad.
\end{equation}

Alternatively, one can write the Bogoliubov transformation as a linear
operator,
\begin{equation}
R = R_1R_2\quad,
\end{equation}
such that
\begin{equation}
\tilde b = \quad R(t)\, b\, R^{\dagger}(t)\quad.\label{eq:btilde}
\end{equation}
$R_1$ and $R_2$ can each be written in terms of only
one of the two commuting SU(2) algebras:
\begin{eqnarray}
R_1 & = & \exp\left[\eta\
I_+\right] \exp\left[\log (1+|\eta|^2)\ I_0\right] \exp\left[-\eta^*\
I_-\right] \nonumber \\
R_2 & = & \exp\left[\eta^*\ T_+\right] \exp\left[\log
(1+|\eta|^2)\ T_0\right] \exp\left[-\eta\
T_-\right]\quad, \label{eq:twoRs}
\end{eqnarray}
where
\begin{equation}
\eta = \frac{-1}{(a^2 - a^{*2})}\,\left[\frac{\tan\theta\,(a\mu_1 A_1 -
a^*\mu_2 A_2)}{\sqrt{k_0^2{\bf A}^2 - ({\bf A\cdot k})^2} }\,\right]
\end{equation}
gives the desired diagonal Hamiltonian.

Writing the Bogoliubov transformation as a linear operator is useful when
calculating the rate of pair production from an electric field.
The physical vacuum after the field has been turned off ($t > T$),
$|Z(T)\rangle$, is related to the vacuum before the field was turned on
($t < -T$), $|0_i\rangle$, by
\begin{equation}
|Z(T)\rangle = R(T)|0_i\rangle\quad.
\end{equation}
Similarly, the physical
creation  operators for fermions and anti-fermions at times $t> T$ are
 $\tilde b^{\dagger}(k)$ and
$\tilde d^{\dagger}(k)$ respectively.\cite{Balant}  The probability amplitude
of producing no pairs, $S_0$, is therefore given by
\begin{eqnarray}
S_0 & = & \lim_{t\rightarrow\infty}\ \langle
Z|U_I|0_i\rangle \nonumber \\
 & = & \langle 0_i|\tilde U|0_i\rangle\quad,
\end{eqnarray}
where
$\tilde U = R^{\dagger}U_I$.  One can
solve for $\tilde U$ directly, through
\begin{displaymath}
i\frac{d\tilde U}{dt}\ = \ \tilde H \tilde U\quad,
\end{displaymath}
where
\begin{equation}
\tilde H\ = \
R^{\dagger}HR - i R^{\dagger}\dot R\quad.
\end{equation}
The probability amplitude
for producing no pairs is related to the rate of pair production,
$\omega$, by
\begin{equation}
|S_0|^2 = e^{-\int\omega\ d^4x}\quad.
\end{equation}

The operator $R_1$ has the matrix representation
\begin{equation}
R_1 =\left(\matrix{\cos\frac{\alpha_1}{2} & \sin\frac{\alpha_1}{2}
e^{-i\gamma_1}\cr
-\sin\frac{\alpha_1}{2}e^{-i\gamma_1} & \cos\frac{\alpha_1}{2}}\right)
\quad,
\label{eq:Rmatrix}
\end{equation}
where $\eta = \tan \frac{\alpha}{2} e^{-i\gamma}$.
$R_2$ is defined analogously.  One observes that, to satisfy
Eq.\ \ref{eq:twoRs} , $\alpha_1 = \alpha_2$ and $\gamma_1 =
-\gamma_2$; the subscripts on $\alpha$ and $\gamma$ are therefore dropped.
The matrix representation allows
one to easily calculate the Hamiltonian $\tilde H$.  By
substituting Eqs.~\ref{eyes} and~\ref{tees} into the Hamiltonian of
Eq.~\ref{eq:tudham}, one can show that this Hamiltonian can be written $H = H_1
+ H_2$, where
\begin{eqnarray}
H_1 & = \int d^3k & \left\{\left(2k_0 - \frac{2eA_ik_i}{k_0}\right)I_0
\right.\nonumber \\
& & + \left. 2ie\mu\ \left[\left(\frac{aA_1}{\mu_2} -
\frac{a^*A_2}{\mu_1}\right)I_+ - \left(\frac{a^*A_1}{\mu_2} -
\frac{aA_2}{\mu_1}\right)I_-\right]\right\}\quad,
\end{eqnarray}
and
\begin{eqnarray}
H_2 & = \int d^3k & \left\{\left(2k_0 - \frac{2eA_ik_i}{k_0}\right)T_0
\right. \nonumber \\
& & + \left. 2ie\mu\ \left[\left(\frac{-a^*A_1}{\mu_2} +
\frac{aA_2}{\mu_1}\right)T_+ + \left(\frac{aA_1}{\mu_2} -
\frac{a^*A_2}{\mu_1}\right)T_-\right]\right\}\quad.
\end{eqnarray}
It
follows that $\tilde H$ can be written $\tilde H = \tilde H_1
+ \tilde H_2$, where
\begin{eqnarray}
\tilde H_1 = R^{\dagger}_1H_1R_1 - iR_1^{\dagger}\dot R_1
\end{eqnarray}
and
\begin{eqnarray}
\tilde H_2 = R^{\dagger}_2H_2R_2 - iR_2^{\dagger}\dot R_2\quad.
\end{eqnarray}
The explicit expressions for $\tilde H_1$ and $\tilde H_2$ are then
\begin{eqnarray}
\tilde H_1 & = & \quad \int d^3k\ \left\{\left[2\varepsilon(k) \, +\,
2(\dot\gamma
\sin^2\frac{\alpha}{2})\right] I_0 \right.\nonumber \\
 &  & \left.\ - \frac{1}{2}\left[ (i \dot\alpha + \dot\gamma
\sin\alpha)e^{-i\gamma} I_+
\ + (-i\dot\alpha\, +\, \dot\gamma \sin\alpha)e^{i\gamma}
I_-\right]\right\}
\label{eq:Htildexp1}
\end{eqnarray}
\begin{eqnarray}
\tilde H_2 & = & \quad \int d^3k\ \left\{\left[2\varepsilon(k) \, -\,
2(\dot\gamma
\sin^2\frac{\alpha}{2})\right] T_0 \right.\nonumber \\
 &  & \left.\ - \frac{1}{2}\left[ (i \dot\alpha - \dot\gamma
\sin\alpha)e^{i\gamma} T_+
\ + (-i\dot\alpha\, -\, \dot\gamma \sin\alpha)e^{-i\gamma}
T_-\right]\right\}\quad.
\label{eq:Htildexp2}
\end{eqnarray}

If we now write
$\tilde U$ as a product, $\tilde U = \tilde U_1 \tilde U_2$, where $\tilde
U_1$ and $\tilde U_2$ are each written in the most general form of an element
of the respective $SU(2)$ groups:
\begin{eqnarray}
\tilde U_1 & = & \exp\left[-i\int
d^3 {\bf k}\ \phi_1 I_0\right] \exp\left[\int d^3{\bf k}\ \tau_1 I_+\right]
\nonumber \\
 &  & \qquad\times\,\exp\left[\int d^3 {\bf k}\ \log
(1+|\tau_1|^2)\,I_0\right]\,\exp\left[-\int d^3 {\bf k}\ \tau_1^* I_-\right]
\end{eqnarray}
and
\begin{eqnarray}
\tilde U_2 & = & \exp\left[-i\int d^3 {\bf
k}\ \phi_2 T_0\right] \exp\left[\int d^3{\bf k}\ \tau_2
T_+\right]\nonumber \\
 &  & \qquad\times\,\exp\left[\int d^3 {\bf k}\ \log
(1+|\tau_2|^2)\,T_0\right]\,\exp\left[-\int d^3 {\bf k}\ \tau_2^* T_-\right],
\end{eqnarray}
then the differential equation for $\tilde U$,
\begin{displaymath}
i \frac{d\tilde U}{dt} = \tilde H \tilde U\quad,
\end{displaymath}
separates into two independent equations
for $\tilde U_1$ and $\tilde U_2$. We show this as follows:
\begin{equation}
i\frac{d\tilde U }{dt} = i\frac{d\tilde U_1}{dt}\tilde U_2 +
i\tilde U_1\frac{d\tilde U_2}{dt} = \left(\tilde H_1 + \tilde
H_2\right)\tilde U_1 \tilde U_2\quad.
\end{equation}
This is the sum of the two equations:
\begin{equation}
\left[i\frac{d\tilde U_1}{dt}= \tilde H_1 \tilde
U_1\right]\tilde U_2
\end{equation}
and
\begin{equation}
\tilde U_1 \left[i\frac{d\tilde U_2 }{dt} = \tilde H_2 \tilde
U_2\right]\quad.
\end{equation}
These are independent differential equations for
$\tilde U_1$ and $\tilde U_2$, each equation containing elements of only one
$SU(2)$ algebra.  When we insert our ansatz for $\tilde U_1$ and $\tilde U_2$
into the above differential equations, we obtain differential equations for the
coefficients $\tau_1$, $\tau_2$, $\phi_1$ and $\phi_2$.  One can proceed to
solve these differential equations in precisely the same manner as in the
one-dimensional case.\cite{Balant}

Let
\begin{equation}
z = \tau \exp(-i\phi +i\gamma)\quad.
\label{eq:zeta}
\end{equation}
The resulting differential equation for $z_1$ is
\begin{eqnarray}
i \dot z_1 & = & \quad -\frac{1}{2}(\dot\gamma\sin\alpha +  i\dot
\alpha) + \nonumber \\
 &  & \quad + 2\left[ \varepsilon(k) + \dot\gamma
(\sin^2\frac{\alpha}{2}+\frac{1}{2})\right]z_1 \label{zeedot1}\\
 &  & \quad + \frac{1}{2}(\dot\gamma\sin\alpha -  i\dot
\alpha) z_1^2\quad. \nonumber
\end{eqnarray}
The corresponding differential
equation for $z_2$ is
\begin{eqnarray}
i \dot z_2 & = & \quad -\frac{1}{2}(-\dot\gamma\sin\alpha +  i\dot
\alpha) + \nonumber \\
 &  & \quad + 2\left[ \varepsilon(k) - \dot\gamma
(\sin^2\frac{\alpha}{2}-\frac{1}{2})\right]z_2 \label{zeedot2}\\
 &  & \quad + \frac{1}{2}(-\dot\gamma\sin\alpha -  i\dot
\alpha) z_2^2\quad. \nonumber
\end{eqnarray}

One can write $\dot \alpha$ and $\dot \gamma$ explicitly, in terms
of the electric field and vector potential.  The expression for
$\dot \alpha$ is:
\begin{equation}
\dot\alpha = \left(\frac{e}{\varepsilon^2}\right) \frac{\left[({\bf k} - e{\bf
A}) \times {\bf E}\right]\cdot ({\bf k}\times{\bf
A})
+ m^2({\bf E}\cdot{\bf A})}
{\sqrt{k_0^2{\bf A}^2 - ({\bf A\cdot k})^2}}\quad,
\end{equation}
with $\dot\alpha = 0$ at $t= -\infty$.
The expression for $\dot\gamma$ is:
\begin{equation}
\dot\gamma =\quad -\frac{\mu k_0 |{\bf A}\times{\bf E}|}{k_0^2{\bf A}^2 -
({\bf A}\cdot k)^2}\quad.
\end{equation}
Note that only $\dot\gamma$ appears explicitly in the equations; when
calculating $|\tau|$ (see Eq.~\ref{eq:szero} below), the initial condition on
$\gamma$ is irrelevant.
When {\bf A} and {\bf E} are parallel ($\dot\gamma = 0$),
Eqs.~\ref{zeedot1} and ~\ref{zeedot2} reduce to the corresponding
equation calculated for the one-dimensional
case.\cite{Balant}  When $\dot\gamma \not= 0$, the two-dimensional nature of
the equations is manifested.

Finally, one can calculate the rate of pair production, $\omega$, via
$S_0$.  The above definitions give us:
\begin{eqnarray}
S_0 = \langle 0_i|\tilde U|0_i\rangle & =\ &
\exp\left[i\int d^3 {\bf k}\,
(\frac{\phi_1 + \phi_2 }{2})\right]\nonumber \\
 &  & \times\ \exp\left[-\int d^3 {\bf k}\,
\log\sqrt{(1 + |\tau_1|^2) (1+|\tau_2|^2)}\right]\quad.
\label{eq:szero}
\end{eqnarray}
Then
\begin{equation}
|S_0|^2 = \ |\langle 0_i| \tilde U|0_i\rangle|^2 =\ e^{-\int d^3 {\bf
k}\,\log [(1+|z_1|^2)(1+|z_2|^2)]}\quad.
\label{eq:Szero}
\end{equation}
Once the differential equations have been solved for $z_1$ and
$z_2$ (in general, this must be done numerically), the
rate of pair production is easily obtained.

To verify this approach, consider the very simple example of a
uniform, static electric field
which is oriented at an angle
$\theta$ to the x-axis.  Then $|{\bf E}\times{\bf A}|=0$, and
$\dot\gamma = 0$.  In this case, $\dot\alpha$ reduces to
\begin{equation}
\dot\alpha = \frac{ek_{\perp} E_0}{\varepsilon^2}\quad,
\end{equation}
where
\begin{equation}
k_{\perp} = \sqrt{k_0^2 - k_{\parallel}^2} = \sqrt{k_0^2 -
\frac{({\bf k\cdot E})^2}{{\bf k}^2}}\quad.
\end{equation}
With these values, the
expressions for $i\dot z_1$ and $i\dot z_2$ are identical, and
each is equal to the expression which applied in the
one-dimensional case.\cite{Balant}  With $z_1 = z_2$,
Eq.~\ref{eq:Szero}
reduces to
\begin{equation}
|S_0|^2 = \ |\langle 0_i| \tilde U|0_i\rangle|^2 =\ e^{-2\int d^3 {\bf
k}\,\log (1+|z_1|^2)}\quad,
\end{equation}
which again is identical to the one-dimensional
result.\cite{Balant}  In this case, the rate of pair production
has been shown to be equal to that calculated by Schwinger, given
in Eq.~\ref{eq:Schwing}.

\section*{III.  CONCLUSIONS}

We have shown that the Hamiltonian describing fermion pair
production from an arbitrarily time-varying electric field in two
dimensions is encompassed by an $SO(4)$ algebra.  We have
also explicitly constructed the two commuting $SU(2)$ algebras in the direct
product $SU(2)\times SU(2)$, which is isomorphic to this $SO(4)$ algebra.  The
one-dimensional problem is described by an $SU(2)$ algebra in the $j=1$
representation, while the two-dimensional problem is described by two $SU(2)$
algebras in the $j=\frac{1}{2}$ representation.  However, when the
one-dimensional problem and the two-dimensional problem are each considered in
the lowest-dimensional representation, one  sees that the off-diagonal elements
are real in the one-dimensional case and complex in the two-dimensional case.
The extra degree of freedom present in the two-dimensional case is manifested
in this way.  Indeed, it can easily be shown that the factor $\gamma$ in
Eq.~\ref{eq:zeta} is the Berry's phase.

This group-theoretic
approach may simplify the calculation of the rate of
fermion pair production from the field, since the two-dimensional
problem can in this way be reduced to two one-dimensional problems.
We verify our approach by showing that the Schwinger formula for
pair production can be obtained for the special case of a
uniform, static electric field.

\bigskip

\section*{Acknowledgments}
This research was supported in part by the U. S. Department of Energy Grant
No. DE-FG02-91ER40652, and in part by the U.S. National Science Foundation
Grant No. PHY-9314131.

\bigskip

\pagebreak

\renewcommand{\baselinestretch}{1}

\end{document}